\documentstyle[12pt]{article}
\newcommand{\wt}{\widetilde}
\newcommand{\wh}{\widehat}

\newcommand{\wca}{\bar}
\newcommand{\hn}{H_N^{(n)}}
\newcommand{\hf}{H_N^{(4)}}
\newcommand{\hfi}{H_N^{(5)}}
\newcommand{\hs}{H_N^{(6)}}
\newcommand{\hepth}[1]{[hep-th/#1]}

\newcommand{\be}{\begin{equation}}
\newcommand{\ee}{\end{equation}}
\newcommand{\ben}{\begin{eqnarray}\displaystyle}
\newcommand{\een}{\end{eqnarray}}
\newcommand{\refb}[1]{(\ref{#1})}

\begin{document}

{}~ \hfill\vbox{\hbox{hep-th/9709220}\hbox{MRI-PHY/P970926}
}\break

\vskip 3.5cm

\centerline{\large \bf D0 Branes on $T^n$ and Matrix Theory}

\vspace*{6.0ex}

\centerline{\large \rm Ashoke Sen
\footnote{E-mail: sen@mri.ernet.in}}

\vspace*{1.5ex}

\centerline{\large \it Mehta Research Institute of Mathematics}
 \centerline{\large \it and Mathematical Physics}

\centerline{\large \it  Chhatnag Road, Jhoosi,
Allahabad 221506, INDIA}

\vspace*{4.5ex}

\centerline {\bf Abstract}

The Hamiltonian describing Matrix theory on $T^n$ 
is identified with
the Hamiltonian describing the dynamics of D0-branes on $T^n$ in
an appropriate weak coupling limit for all
$n$ up to 5. New subtleties arise in taking this
weak coupling limit for $n=6$, since the transverse size of the D0 brane
system blows up in this limit.
This can be attributed to the appearance of extra light
states in the theory from wrapped D6 branes. This subtlety is
related to the difficulty in finding a Matrix formulation of
M-theory on $T^6$.

\vfill \eject

\baselineskip=18pt

During the last year great progress has been made in finding a
non-perturbative formulation of 
$M$-theory\cite{BFSS,SUSS,BECKER,BBPT}
and its toroidal compactification\cite{BFSS}-\cite{SEIBERG}. This
formulation, known as Matrix theory, can be summarized into the
following recipe:

\begin{itemize}
\item
Begin with type IIA string theory on 
an $n$ dimensional torus $T^n$. 
For simplicity we shall take the torus to be rectangular,
without any background anti-symmetric tensor fields, but these
can easily be introduced. Let $m_S^2$ be the string tension, 
$R_i$ ($1\le i\le n$)
be the radii of the $n$ circles on $T^n$ (measured in
physical units), and $g_S$ be the string coupling constant. By using
the usual duality between $M$-theory on $S^1$ and type IIA string
theory\cite{TOWNSEND,WITTEN}, this theory can be regarded as
$M$-theory on $S^1\times T^n$. The radius $R$ of $S^1$, and the
eleven dimensional Planck mass $m_p$ are related to $g_S$ and
$m_S$ through the relations:
\be \label{e1}
R=g_S/m_S, \qquad m_p=m_S/g_S^{1/3}\, .
\ee

\item We consider the dynamics of $N$ D0 branes in this type IIA
string theory, and consider the weak coupling limit $g_S\to 0$ keeping
the following quantities fixed:
\be \label{e2}
r_i \equiv m_p R_i = g_S^{-1/3} m_S R_i, \qquad M\equiv m_p^2 R
=m_S g_S^{1/3}\, .
\ee
Keeping $r_i$ fixed corrresponds to keeping the size of $T^n$,
measured in units of $m_p^{-1}$, fixed. As we shall see, keeping 
$M$ fixed guarantees that we keep the `correct' degrees of freedom
$-$ necessary for obtaining the Matrix theory Hamiltonian $-$
on the D0 brane world volume, throwing away the rest. 
In this limit, the Hamiltonian
describing the dynamics of $N$ D0 branes can be parametrized by
the parameters $M$ and $r_i$. Let us denote this by 
\be \label{enew1}
\hn(M, \{r_i\}).
\ee

\item 
Let us now consider M-theory on $T^n$, with radii $L_i$.
According to the Matrix theory conjecture as stated in \cite{SUSS}, 
the Discrete Light
Cone Quantization (DLCQ) of this theory, in which we
compactify another light-like direction on a circle of radius
$L$, and study the sector carrying N units of momentum in the
light like direction, is described {\it exactly} by the
Hamiltonian:
\be \label{e3}
\hn(M_p^2 L, \{M_pL_i\}).
\ee
Here $M_p$ is the Planck mass of this new $M$-theory whose
matrix description we are seeking.
This prescription allows us to have a complete non-perturbative
formulation of $M$-theory in terms of weak 
coupling dynamics of $D0$ branes in type IIA string theory.

\end{itemize}

Since this recipe appears to differ from the conventional recipe
for the Matrix formulation of $M$-theory on $T^n$, we shall first
show that this is equivalent to the conventional recipe
for all $n$ up to 5. We shall then examine the difficulty in
extending this recipe to give a Matrix formulation of $M$-theory
on $T^6$\cite{LMS}-\cite{GANOR}.
In showing the equivalence of this recipe with the conventional
recipe, we shall make use of various known U-duality
transformations in type II string theory to map the dynamics of
D0 branes in type IIA on $T^n$ to the dynamics of various other
systems. Let us call the definition of $\hn$ that we have given
as description $A$. In this description the parameters $m_S$ and $R_i$
are related to the finite parameters $M$ and $r_i$ as follows:
\be \label{e4}
m_S = M g_S^{-1/3}, \qquad R_i = M^{-1} r_i g_S^{2/3}\, .
\ee

We shall now give a second description of $\hn$, which we shall
call description B, by making an $R\to 1/R$ duality
transformation in all the $n$ directions on $T^n$. This converts
the type IIA theory to type IIA/IIB theory on a dual torus
$\wt T^n$ depending on whether
$n$ is even or odd, and maps the system of $N$ D0 branes to a
system of $N$ D$n$-branes wrapped on $\wt T^n$. The string mass
$\wt m_S$, the radii $\wt R_i$ of $\wt T^n$, and the string coupling
$\wt g_S$ in this new theory are given by:
\ben \label{e5}
&& \wt m_S = m_S = M g_S^{-1/3}, \qquad \wt R_i = m_S^{-2}R_i^{-1}
=M^{-1} r_i^{-1}, \nonumber \\
&& \wt g_S = g_S/\prod_{i=1}^n(m_S R_i)
=g_S^{1-{n\over 3}}\Big(\prod_{i=1}^n r_i^{-1}\Big)\, .
\een
Thus $\hn$ can also be regarded as the Hamiltonian describing the
dynamics of $N$ wrapped D$n$ branes in this theory in
the $g_S\to 0$ limit.

For $n=5$, we shall give yet
another description of $\hn$, which we shall call description C,
by making an S-duality
transformation in the type IIB string theory\cite{HT}. This
converts the wrapped D5-branes to wrapped NS five branes, and
gives the following new set of parameters labelling string mass
scale $\wh m_S$, radii $\wh R_i$ of the five torus
$\wh T^5$, and the
string coupling $\wh g_S$ respectively:
\ben \label{e6}
&& \wh m_S = \wt m_S \wt g_S^{-1/2}=M\Big(\prod_{i=1}^5
r_i^{1/2}\Big),
\qquad \wh R_i = \wt R_i =M^{-1} r_i^{-1},  \nonumber \\
&& \qquad \wh g_S = \wt g_S^{-1}
=g_S^{2\over 3}\Big(\prod_{i=1}^5 r_i\Big)\, .
\een

At this stage we are ready to compare $\hn$ as defined above with
the conventional description of Matrix theory on $T^n$ for $n\le
5$. First of all, note that for $n=5$, as we take the $g_S\to 0$
limit in description C, the new string coupling $\wh g_S$
vanishes, with $\wh m_S$ and $\wh R_i$ approaching finite value.
Thus
$\hfi$ corresponds to the Hamiltonian describing the dynamics of
wrapped NS five branes in type IIB on $T^5$ in the limit of zero
string 
coupling. This is precisely the Matrix formulation of $M$-theory
on $T^5$ as proposed in \cite{SEIBERG}. The relationship
between the parameters of this new IIB theory, and the original
variables also work out correctly if we identify \refb{e3} as the
Matrix theory Hamiltonian. Incidentally, this analysis
can also be repeated for D0 branes moving on $K3\times S^1$
instead of $T^5$, and the weak coupling limit of this theory
correctly reproduces the proposed Matrix description of
$M$-theory on $K3\times S^1$\cite{SURESH,BERKOOZ}. 
The only difference is
that in going from the description A to description B,
making $R\to (1/R)$ duality on the first four directions
need to be replaced by making a mirror transformation on $K3$.

Once we have found agreement for $n=5$, the agreement for all
other $n\le 4$ is guaranteed, since these can be obtained from
the $n=5$ Hamiltonian by taking one or more $r_i$'s to infinity.
However, one can also verify this explicitly. For example, for
$n\le 3$ we can use the description B and take the $g_S\to 0$
limit. In this limit, $\wt m_S$ approaches infinity
and $\wt g_S$ either approaches
0 (for $n\le 2$) or remains finite (for $n=3$). 
$\wt R_i$ remain finite. Thus we can safely ignore the effect
of the massive string modes, as well as higher derivative terms in the
effective action, and  the effective dynamics of the
wrapped $Dn$-branes is described by $(n+1)$
dimensional supersymmetric $U(N)$ gauge theory compactified on
$\wt T^n$. The bosonic part of
this action is given by:
\ben  \label{e7}
&& {\wt m_S^{n-3}\over \wt g_S}\int dt \int d^nx
\Big( Tr (F_{\mu\nu}F^{\mu\nu}) + \sum_{\alpha=1}^{9-n}
Tr(D_\mu \Phi^\alpha D^\mu \Phi^\alpha) + 
\sum_{\alpha,\beta} Tr\big([\Phi^\alpha, 
\Phi^\beta]^2\big)\Big) \nonumber \\
=
&& M^{n-3} \Big(\prod_{i=1}^n r_i\Big) \int dt \int d^nx
\Big( Tr (F_{\mu\nu}F^{\mu\nu}) + \sum_{\alpha=1}^{9-n}
Tr(D_\mu \Phi^\alpha D^\mu \Phi^\alpha) + 
\sum_{\alpha,\beta} Tr\big([\Phi^\alpha, \Phi^\beta]^2\big)\Big) \,
, \nonumber \\
\een
where $\Phi^\alpha$ are scalar fields in the 
adjoint of $U(N)$, normalized so
as to have mass dimension unity. This is
precisely the proposed Hamiltonian for Matrix theory on $T^n$
for $n\le 3$.

In order to check explicitly that the recipe also correctly
reproduces the Hamiltonian for Matrix theory on $T^4$, we 
start with the description B, and regard this as $M$-theory
compactified on $\wt S^1\times \wt T^4$. The wrapped D4 branes
can then be regarded as $M$-theory five branes wrapped on $\wt
S^1\times \wt T^4$. We shall call this description D. The radius
$\wt R$ of $\wt S^1$ and the new Planck mass $\wt m_p$ are given
by, 
\be \label{e8}
\wt R = \wt g_S/\wt m_S = M^{-1} \Big(\prod_{i=1}^4
r_i^{-1}\Big), \qquad
\wt m_p = \wt m_S/\wt g_S^{1/3} = M g_S^{-2/9} \Big(\prod_{i=1}^4
r_i^{1/3}\Big)\, .
\ee
Thus in the $g_S\to 0$ limit $\wt R$ and $\wt R_i$ remain finite,
and $\wt m_p$ approaches $\infty$. $\hf$ is given by the
Hamiltonian of wrapped five branes in this limit. This is
precisely the Matrix theory Hamiltonian for $M$-theory on $T^4$
as proposed in \cite{ROZALI,BRS,SEIBERG}.

The recipe for constructing $\hn$ given here
can be applied to any
compactification $-$ toroidal or otherwise $-$
although whether it gives a sensible and correct matrix description
of 
$M$-theory is an altogether different question. We shall now try
to apply this to construct $\hn$ for $n=6$. This can be done by
compactifying one of the non-compact directions on a circle of
radius $R_6$, and taking the $g_S\to 0$ limit keeping fixed
\be \label{e9}
r_6 = m_p R_6 = M g_S^{-2/3} R_6.
\ee
We can use any of the four descriptions to do this; but it
will be most convenient to start with the description $C$. Thus we
have type IIB on a torus $\wh T^6$ with parameters as given in
\refb{e6}, and the radius of the sixth circle given by
\be \label{e10}
\wh R_6 = R_6 = M^{-1} g_S^{2/3} r_6\, .
\ee
$\hs$ corresponds to the Hamiltonian
of NS five branes wrapped on 1-5
directions in the $g_S\to 0$ limit. In order to take this limit,
it will be convenient to go to a new description of $\hs$ by
making an $R\to (1/R)$ duality in the 6th direction. This
converts type IIB to type IIA and the NS five branes to
Kaluza-Klein monopoles associated with the sixth direction.
The parameters in this theory are:
\ben \label{e11}
&& \wca m_S = \wh m_S =M\Big(\prod_{i=1}^5 r_i^{1/2}\Big),
\qquad   \wca R_i = \wh R_i =M^{-1} r_i^{-1} \quad
\hbox{for} \quad 1\le i\le 5,  \nonumber \\
&& \wca R_6 = \wh m_S^{-2} \wh R_6^{-1} = M^{-1} 
\Big(\prod_{i=1}^5 r_i^{-1}\Big) r_6^{-1} g_S^{-2/3}, \nonumber \\
&& \wca g_S = \wh g_S \wh m_S^{-1} \wh R_6^{-1}
=\Big(\prod_{i=1}^5 r_i^{1/2}\Big) r_6^{-1} \, .
\een
We shall call this description E. These Kaluza-Klein monopoles
can also be described as type IIA on $\wca T^5\times MTN\times R$, where
$\wca T^5$ denotes the torus labelled by 1-5 directions, 
$MTN$ is the multi-Taub-NUT space and $R$ denotes the usual time
direction. 

We now take the $g_S\to 0$ limit. Note that in this limit the
radii of $\wca T^5$, $\wca g_S$ and $\wca m_S$ remain
finite. Furthermore, the dimensionless parameters
$\wca m_S\wca R_i$
for $(1\le i\le 5)$ and $\wca g_S$  are all independent. 
Thus this theory is
expected to have the full U-duality symmetry of type IIB on $\wca
T^5$, and it is easy to see that indeed, when we put in the
appropriate 3-form field background and the general flat metric on
$T^6$ in the original description
A, the moduli
space of this theory has the structure of $E_{6(6)}(Z)\backslash
E_{6(6)}(R)/Sp(4)$, which is the expected structure of the moduli
space for $M$-theory on $T^6$. However, note that in the
$g_S\to 0$ limit $\wca R_6\to\infty$. Since $\wca R_6$ sets the
overall size of the Taub-NUT space, we see that in this limit the
Taub-NUT space expands to infinite size! 
In other words, the Taub-NUT space becomes the ALE space with 
$A_{N-1}$ singularity\cite{AGH}. (This problem is 
clearly related to the result 
of \cite{SEISET}). For $N=1$, we just get back four 
dimensional Euclidean space $R^4$.  Since the
transverse space is now non-compact, we effectively get a (4+1) 
dimensional theory instead  of a (0+1) dimensional theory. We
shall not address the question as to whether this theory can in
any way be useful in finding a Matrix formulation of
$M$-theory on $T^6$. Instead, we shall try to analyze the reason
behind getting a (4+1) dimensional theory in the first place. 

Clearly the reason behind getting a (4+1) dimensional theory is
the increase in the transverse size of the system, so we shall
focus on this problem. For this we go back to the description A,
and try to identify the various BPS states in the bulk theory. If
any of them becomes light, then the interaction of these states
with the D0-brane system might be responsible for the increase in
the size of the system. The possible BPS states are as follows:

\begin{enumerate}

\item
States carrying Kaluza-Klein momentum in the transverse
direction. These states have masses of order:
\be \label{e12}
R_i^{-1} = M g_S^{-2/3} r_i^{-1} \to \infty \quad {\rm as}
\quad g_S\to 0.
\ee
Thus these states disappear from the spectrum in the
$g_S\to 0$ limit.

\item
String winding modes. They have mass of order
\be \label{e13}
m_S^2 R_i = Mr_i\, .
\ee
These masses remain finite in the $g_S\to 0$ limit.

\item
For $n\ge 5$, we can have NS five-branes wrapped on $T^5$. A
five-brane wrapped in the first five directions has a
mass of order:
\be \label{e14}
g_S^{-2} m_S^6 \Big(\prod_{i=1}^5 R_i\Big) = g_S^{-2/3} M 
\Big(\prod_{i=1}^5
r_i\Big)\, \to \infty \quad \hbox{as} \quad g_S\to 0.
\ee
Thus these states also disappear from the spectrum as $g_S\to 0$.

\item
Dirichlet $p$ branes wrapped on a $p$-cycle. Of course since we
are dealing with type IIA theory, $p$ must be even. Also a given
$p$-brane can appear only for $n\ge p$. Such a $p$-brane, wrapped
on the first $p$ cycles of $T^n$ has mass of order:
\be \label{e15}
g_S^{-1} m_S^{p+1} \Big(\prod_{i=1}^p R_i\Big) = g_S^{(p-4)/3} M
\Big(\prod_{i=1}^p r_i\Big)\, .
\ee
For $p<4$, these states have infinite mass in the $g_S\to 0$
limit, and hence decouples from the theory. For $p=4$ they have
finite mass, representing new degrees of freedom from the bulk that
might interact with the D0-brane system.
Indeed, they form marginally stable bound state with the D0-brane
system thereby opening up a whole new 
dimension\cite{ROZALI,BRS}. Finally, for $p=6$, these
states become massless, opening up the possibility that
interaction of these states with the D0-brane system can
effectively increase the transverse size of the system. Indeed,
for finite but small $g_S$, these states have mass of order
$g_S^{-2/3}$, precisely the inverse of the transverse size of the
Multi-Taub-NUT space that we had seen in eq.\refb{e11}.

\end{enumerate}

One can make this more concrete by studying what a wrapped D6
brane corresponds to in description E. The series of duality
transformations that leads us from the description A to
description E transforms a wrapped D6 brane in description A to a
state carrying Kaluza-Klein momentum in the 6th direction in
description $E$. Let us now recall the reason as to why the
transverse size of the Taub-NUT solution is related to the size
$\wca R_6$ of the 6th direction. One can write down a solution of
Einstein's equation, in which the transverse size is independent
of $\wca R_6$; however, this solution will suffer from conical
singularities unless the transverse size matches $\wca R_6$. These
conical singularities can only be sensed by states carrying
momentum along the 6th direction, since a state carrying no
momentum in this direction will not sense the periodicity in this
direction, and hence will not see the sigularity. Thus we see
that it is indeed the interaction of the Taub-NUT space with the
states carrying momentum in the 6th direction that is responsible
for the large transverse size of the TN space.

It is now time to summarize our results. By adopting a uniform
approach to the Hamiltonian for Matrix theory on $T^n$
in terms of weak coupling dynamics of D0 branes on $T^n$, we can
give a naive description of this Hamiltonian for any $n$. We have
shown that this naive description produces a (4+1) dimensional
Hamiltonian instead of a (0+1) dimensional Hamiltonian, $-$ the
Hamiltonian describing type IIB string theory on a dual torus
$\wca T^5$ times the ALE space, at finite coupling. 
This clearly has the required U-duality symmetry $E_{6(6)}(Z)$
that is expected of a Matrix theory on $T^6$; however how a (4+1)
dimensional theory could be useful in constructing a Matrix
theory remains unclear.
The reason
for getting a (4+1) dimensional theory can be traced to the
increase of the transverse size of the D0-brane system in the
weak coupling limit. This, in turn, is due to the interaction of
the D0-branes with wrapped D6-branes, which become massless in
the weak coupling limit we are considering. 

Note added: After the paper was sent to the archieve, another
paper\cite{SEINEW} containing very similar results appeared. This
paper also provides an explanation of why this limit gives the
correct quantum mechanical
model for Matrix theory. It has been pointed out by M.~Douglas
that when the compact space is curved, there might be new
subtleties due to the issues raised in \cite{DOS,GRDO}. Finally,
I wish to thank L.~Motl for his comments on the manuscript.

\end{document}